\documentclass[twocolumn,aps,prl,superscriptaddress]{revtex4-2}

\usepackage{amsmath}
\usepackage{graphicx}
\usepackage{MnSymbol}
\usepackage{parskip}
\usepackage{makecell}
\usepackage{booktabs}
\usepackage{multirow}

\begin{document}

\title{Plastic smoothing of rough surfaces}
\author{B.N.J. Persson}
\affiliation{State Key Laboratory of Solid Lubrication, Lanzhou Institute of Chemical Physics, Chinese Academy of Sciences, 730000 Lanzhou, China}
\affiliation{Peter Gr\"unberg Institute (PGI-1), Forschungszentrum J\"ulich, 52425, J\"ulich, Germany}
\affiliation{MultiscaleConsulting, Wolfshovener str. 2, 52428 J\"ulich, Germany}

\begin{abstract}
When two metal blocks are squeezed together the stresses in the asperity contact regions are usually so large that
the asperities deform plastically, at least at short length scales. 
Many tribology properties of contacts, such as the contact stiffness and the electric and thermal contact resistance, and the fluid flow at interfaces,
depend on the surface topography and are hence modified by the plastic flow. Here I present a new way to obtain an effective power spectra of
plastically deformed surfaces to be used in the Persson contact mechanics theory. I also present results for the
surface height topography obtained using the plastically modified power spectra,
and compare to the experimental results of Yusof and Ripin, who studied the influence of plastic flow on the 
topography for a smooth steel surface squeezed against a rough steel surface. Finally, I discuss 
why some surfaces after plastic deformation have similar Gaussian roughness as before plastic deformation, only with smaller roughness amplitude,
while other surfaces shows very skewed roughness after plastic deformation.
\end{abstract}

\maketitle

\setcounter{page}{1}
\pagenumbering{arabic}




{\bf 1 Introduction} 

All surfaces have roughness on many length scales \cite{Weber,Ali,Brod} which affect almost all tribology problems. 
When metallic bodies are squeezed together the local stress in asperity contact regions 
will almost always become so large that the solids deform plastically. Thus, even if no plastic flow occur at the macroscale (the scale of the object
under study, or the width of the nominal contact region), at some shorter length scales plastic flow will almost occur in applications involving
metals, e.g., metallic seals \cite{Metal}. Many experiments have been performed to study how the plastic flow affect the surface roughness. Such studies are not trivial since the topography
need to be studied along exactly the same line (or the same surface area) before and after the plastic contact. 

\begin{figure}
\includegraphics[width=0.35\textwidth,angle=0.0]{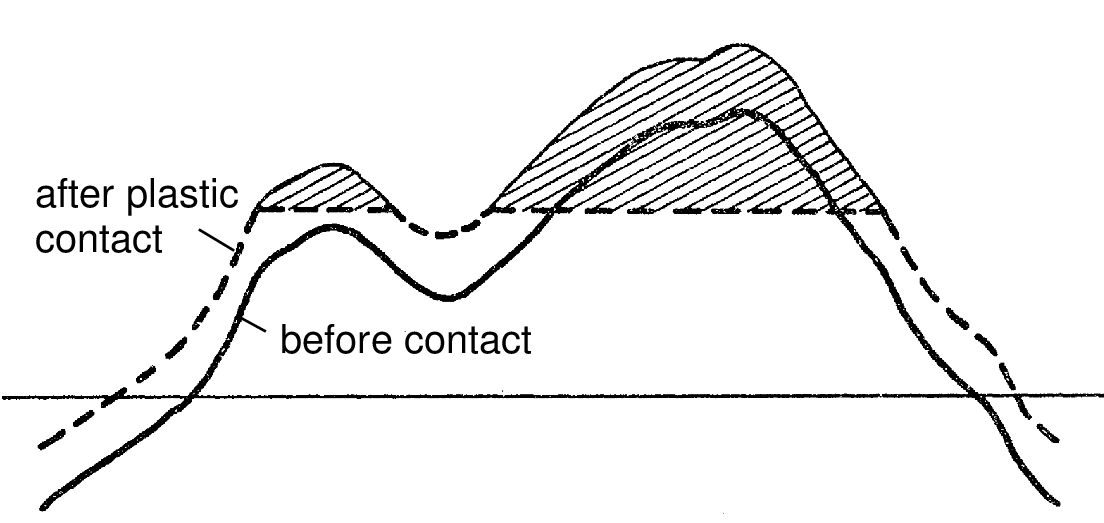}
\caption{\label{old.eps}
The solid line is the asperity profile before contact with a flat rigid surface, and the dashed line after plastic contact. The dashed
area result in a uniform rise in such a way that the volume of the solid is conserved. Adapted from \cite{Williams}.
}
\end{figure}

Theories have been presented for how plastic flow change the topography of rough surfaces. Since plastic flow gives rise to negligible change in the metal volume
(a small change may occur due to residual elastic strain, or due to the creation of defects) an accurate theory must conserve the volume of the metal. 
Since plastic flow is a complex process, in particular
for surfaces with roughness on many length scales (as is always the case), simple ideas have been presented for for plastic flow.
Thus, Pullen and Williamson\cite{Williams} assumed that the material removed from the top of asperities gives rise to a uniform rise
as indicated in Fig. \ref{old.eps}. Here the solid removed (dashed area) lift the profile upwards (dashed line) by a distance determined by the condition that the
volume of metal does not change. However, the experiments of Yusof and Ripin \cite{exp} reviewed below show that the assumption of a uniform rise is not accurate, at least for the
contact pressures used in their study.

The numerically most effective method to study the elastic contact between surfaces with multiscale roughness is the boundary element method (BEM) \cite{Mus,Alm} where the
bulk degrees of freedom is integrated out resulting in a 2D problem rather than a 3D problem. The BEM model has been extended to include plasticity by assuming that
when the normal stress on a surface grid patch (a square area $a^2$ surrounding a grid point, where $a$ is the separation between two nearby grid points) 
becomes larger than the penetration hardness the grid point moves downwards so that the stress equals the penetration hardness. However, this procedure does not conserve the volume
of the solid. Another approach to the contact between randomly rough surfaces is the Persson contact mechanics theory \cite{ep3,ep4,ep5} (see also \cite{ep1,ep2}), 
which has also been used for surfaces which can deform
plastically. In this approach one first calculate the surface roughness power spectrum of the plastically deformed surface by reducing the amplitude of the roughness
wavelength components which deform plastically. This procedure conserve the volume of the solid and, as will be shown below, effectively correspond to
moving the material removed from the top of asperities towards the valleys.

Including the volume conservation condition is very important in all applications where the separation between the surfaces is important, e.g., for the contact stiffness
(which equals $K= -d\sigma_0/d\bar u$, where $\sigma_0$ is the nominal (applied) pressure and $\bar u$ the average surface separation), or for the fluid flow at interfaces, e.g., leakage of
metallic seals. Thus, the plastic flow from the top of asperities to the valleys \cite{explain} will reduce the average surface separation $\bar u$, which will increase the contact stiffness
and increase the viscous resistance to fluid flows at interfaces.

In this work I will compare the prediction of the Persson contact mechanics theory including plasticity with experimental data of Yusof and Ripin. The comparison is only qualitative as
the power spectra of the surfaces used in Ref. \cite{exp} was not given. However, both the theory and the experiments show that the material removed by plastic flow ends up mainly in
the valleys and line scans of the measured and calculated surface topographies are statistically very similar. The experiments by Yusof and Ripin focused on the case where
the nominal contact pressure $p_0$ was such the relative (plastic) contact area was $A/A_0 \sim 0.1$. Other studied have shown
that for much higher contact pressures the contact area approach $\approx 0.5$. The fact that the contact area saturates at $\approx 0.5$ as $p_0 \rightarrow \infty$
has been attributed to plasticity mechanics of asperity interaction. Qualitatively, when an asperity becomes strongly plastically deformed the stress field 
approach a hydrostatic stress and the asperity therefore becomes resistant to further plastic deformation.
We are not aware of any theoretical study which includes this type of plasticity mechanics for surfaces with multiscale roughness .

Since the plastic removed material ends up mainly in the valleys, for not too high contact pressures the area of real contact, the contact morphology, 
and the contact stress distribution, may be well described by the simple approach
used in the BEM studies where grid points are moved vertically so that the contact stress is capped by the penetration hardness $\sigma_{\rm P}$. 
However, physical quantities which depends on the gap between the solids will not be correctly described by this pressure-cap approximation.

In a recent paper it has been shown that the Persson contact mechanics theory with plasticity gives nearly the same elastic and plastic contact area as the
BEM method\cite{Alm1}, but the contact stiffness was higher than predicted by the BEM method\cite{Alm2}. 
This may be related to the flow of solids from the top of asperities to the valleys.

\begin{figure}
\includegraphics[width=0.25\textwidth,angle=0.0]{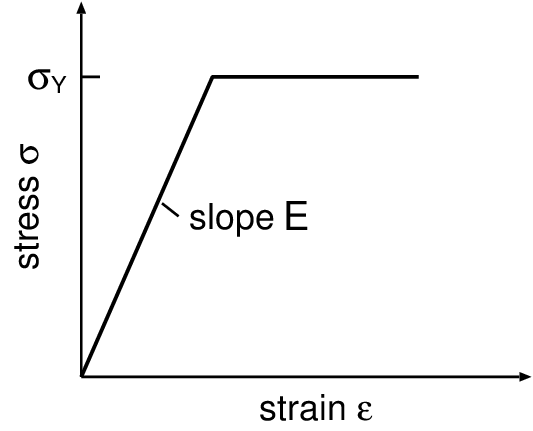}
\caption{\label{ElastoPlasticStressStrain.eps}
The relation between stress and strain in uniaxial tension for the simplest elastoplastic material model. The material deforms elastically with slope $E$ up to the yield stress $\sigma_{\rm Y}$, beyond which it flows plastically without strain hardening.
}
\end{figure}

\vskip 0.3cm
{\bf 2 Multiscale contact mechanics theory} 

We briefly review the Persson contact mechanics theory including plasticity.
We study the interface between an elastoplastic solid with a randomly rough surface, and a rigid body with a perfectly flat surface at a magnification $\zeta$. 
At this magnification, we only observe the roughness with wavenumber components below $\zeta q_0$, where $q_0$ is the wavenumber of the smallest roughness component. 
We assume that the solids have nominally flat surfaces and are squeezed together with a nominally uniform stress. The nominal contact area is denoted by $A_0$. 
Because of the surface roughness, the stress acting at the interface is highly non-uniform. 
In the Persson contact mechanics theory for elastic solids, the probability distribution of contact stress equals \cite{ep3,ep4,ep5}:
\begin{equation}\label{eq1}
    P(\sigma,\zeta) = \dfrac{1}{(4\pi G)^{1/2}} \left (e^{-(\sigma-\sigma_0)^2/4G}- e^{-(\sigma+\sigma_0)^2/4G} \right ),
\end{equation}
where $\sigma_0$ is the nominal (applied) pressure, and where
\begin{equation}\label{eq2}
    G (\zeta) =\frac{\pi}{4} (E^*)^2 \int_{q_0}^{\zeta q_0} dq \ q^3 C(q)S(q),
\end{equation}
where $E^* = E/(1-\nu^2)$ is an effective elastic modulus and $C(q)$ the surface roughness power spectrum.

The factor $S(q)$ in (2) takes into account that the elastic energy stored in the contact regions is smaller than what would be the case if complete contact would occur
in the contact regions observed at the magnification $\zeta$.
When the relative contact area $A/A_0 << 1$, as in the applications below, $S(q)\approx\gamma$ (see Ref. \cite{ep11}). 
The elastic energy reduction factor $\gamma$ has been determined originally by comparing the average surface separation as a function of squeezing pressure with numerical simulation results, and has been found to be (see Ref. \cite{ep11}) $\gamma\approx 0.5$. 
Note that using the correction factor $S(q)\approx\gamma$ is equivalent to using $S=1$ and a reduced modulus $E_{\rm eff} \approx E \surd \gamma \approx 0.73 E$.

When the stress in the asperity contact region becomes high enough, plastic flow occurs. In the simplest model, it is assumed that a material deforms as a linear elastic solid until the stress reaches a critical level, the so-called plastic yield stress, where it flows without strain hardening, see Fig.~\ref{ElastoPlasticStressStrain.eps}. The yield stress in elongation is denoted by $\sigma_{\rm Y}$. In indentation experiments, where a sharp tip or a sphere is pushed against a flat solid surface, the penetration hardness $\sigma_{\rm P}$ is defined as the ratio between the normal force and the projected (on the surface plane) area of the plastically deformed indentation. As shown by Tabor\cite{Tabor}, typically $\sigma_{\rm P} \approx 3 \sigma_{\rm Y}$. We note that the yield stress of materials often depends on the length scale (or magnification), which in principle can be included in the formalism we use \cite{Preview,Brodsky}.

The influence of plastic flow on the contact mechanics is taken into account in the Persson contact mechanics approach by replacing the 
boundary condition $P(\infty,\zeta) = 0$ with $P(\sigma_{\rm Y},\zeta) = 0$. This approach is based on the simplest elastoplastic description, 
where only elastic deformation occurs for $\sigma < \sigma_{\rm P}$, while for $\sigma = \sigma_{\rm P}$, 
the material flows without work-hardening so that the maximal stress equals $\sigma_{\rm P}$ (see Fig.~\ref{ElastoPlasticStressStrain.eps}).

The pressure probability distribution in the elastoplastic theory, in the region where elastic deformation has occurred, is given by~\cite{ep4}:
\begin{equation}\label{eq3}
    P(\sigma,\zeta) = \frac{2}{\sigma_{\rm P}} \sum_{n=1}^\infty \, {\rm sin} (s_n\sigma_0) \, {\rm sin} (s_n\sigma) \, e^{-s_n^2 G(\zeta)},
\end{equation}
where $s_n = n \pi/\sigma_{\rm P}$. 
Note that, as $\sigma_{\rm P} \rightarrow \infty$, \eqref{eq3} reduces to \eqref{eq1}. 
If $A_{\rm el} (\zeta) /A_0$ is the fraction of the nominal area where the solids are in elastic contact when the interface is observed at the magnification $\zeta$, then 
\begin{align}
    &\frac{A_{\rm el} (\zeta )}{A_0} = \int_0^{\sigma_{\rm P}} d \sigma \ P(\sigma,\zeta) \nonumber \\[2ex]
    & = \frac{2}{\pi} \sum_{n=1}^\infty \, \frac{1}{n} \left[ 1-(-1)^n\right ] {\rm sin} (s_n\sigma_0) \, e^{-s_n^2 G(\zeta)} \label{eq4}
\end{align}
The fraction of the nominal area where the solids are in plastic contact when the interface is observed at the magnification $\zeta$ is
\begin{equation}\label{eq5}
    \frac{A_{\rm pl} (\zeta )}{A_0} = \frac{\sigma_0}{\sigma_{\rm P}}+\frac{2}{\pi} \sum_{n=1}^\infty \, \frac{(-1)^n}{n} {\rm sin} (s_n\sigma_0) \, e^{-s_n^2 G(\zeta)}.
\end{equation}

\begin{figure}
\includegraphics[width=0.48\textwidth,angle=0.0]{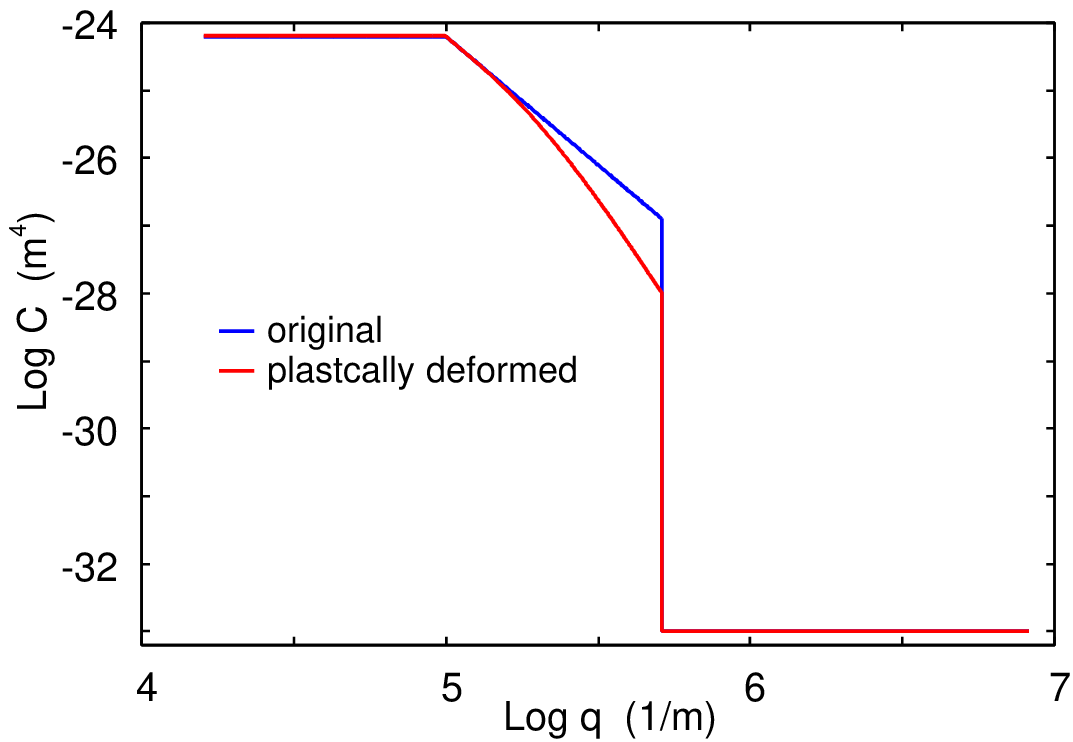}
\caption{\label{1logq.2logC.plastic.eps}
The original surface roughness power spectrum $C(q)$ (blue line) and after plastic deformation
$C_{\rm pl}(q)$ (red line) as a function of the wavenumber $q$ (Log-Log scale).
$C_{\rm pl}(q)$ is obtained from (10) with $E=200 \ {\rm GPa}$, $\nu = 0.3$, $\sigma_{\rm P} = 3.2 \ {\rm GPa}$
and $\sigma_0= 0.2 \ {\rm GPa}$.
}
\end{figure}

\begin{figure}
\includegraphics[width=0.48\textwidth,angle=0.0]{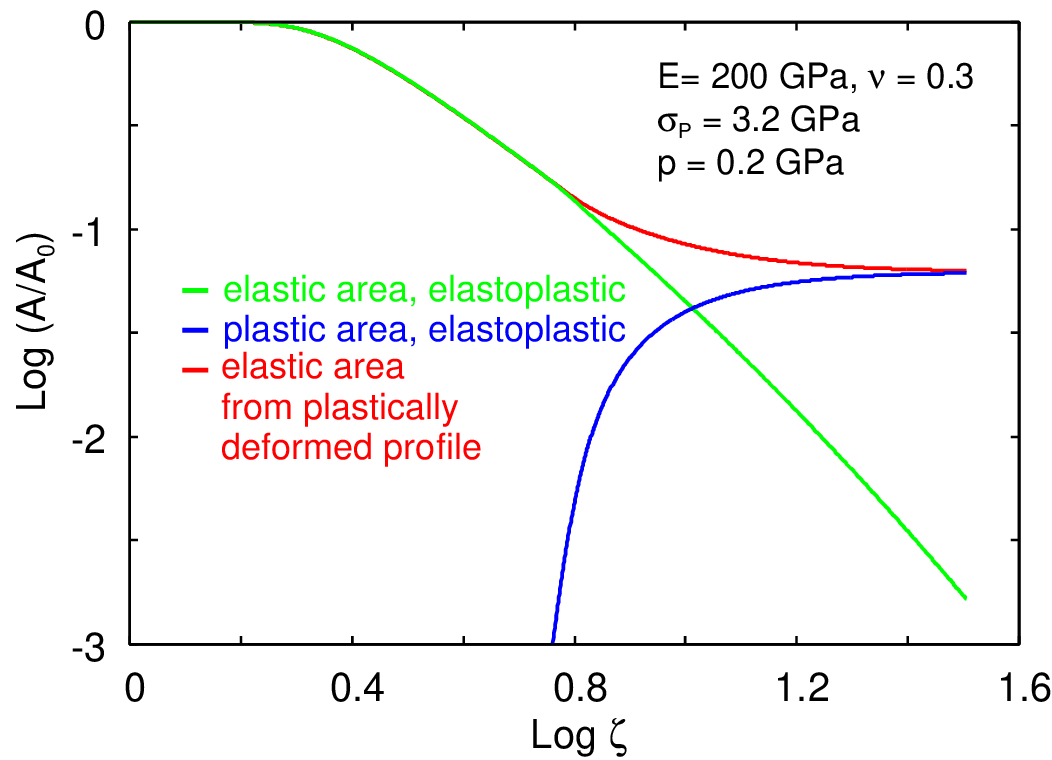}
\caption{\label{1logZeta.2LogArea.all.eps}
The elastic contact area $A_{\rm el}$ (green line) and the plastic contact area $A_{\rm pl}$ (blue line) as a function
of the wavenumber $q$ (Log-Log scale). The red line is the contact area calculated assuming elastic contact with 
the power spectrum $C_{\rm pl}$ [given by (10)] of the plastically deformed surface profile.
}
\end{figure}

We will illustrate the plastic contact mechanics theory for a system similar to that studied experimentally by
Yusof and Ripin. We consider squeezing a rigid and flat surface against an elastoplastic solid with surface roughness.
The surface roughness power spectrum of the surface is shown by the blue curve in Fig. \ref{1logq.2logC.plastic.eps} and has the
root-mean-square roughness $0.2 \ {\rm \mu m}$ and the slope of the tilted line $-3.8$ correspond to a self-affine surface with the Hurst exponent $H=0.9$.
The roll-off and cut-off wavenumbers, $q_{\rm r}$ and $q_{\rm c}$, are chosen to be similar to in the experiments of Yusof and Ripin. Thus, the cut-off 
corresponding to the wavelength $\lambda_{\rm c} = 2 \pi /q_{\rm c} \approx 12.5 \ {\rm \mu m}$ which reflect the resolution of the 
optical method they used to study the surface, below which they cannot detect surface roughness.

Fig. \ref{1logZeta.2LogArea.all.eps}
shows the elastic contact area $A_{\rm el}$ (green line) and the plastic contact area $A_{\rm pl}$ (blue line) as a function
of the wavenumber $q$ (Log-Log scale). 
The results are for an elastic solid with the Young's modulus $E=200 \ {\rm GPa}$, the Poisson ratio $\nu = 0.3$ 
squeezed against the rigid flat solid with the nominal pressure $\sigma_0= 0.2 \ {\rm GPa}$.
We have used the steel penetration hardness $\sigma_{\rm P} = 3.2 \ {\rm GPa}$ which is typical for the work-hardened steel
used in the experiments.
(The experiments used 310 stainless steel which has the penetration hardness $\approx 2 \ {\rm GPa}$, which increases
to $2.5-3.5$ when work hardened. We assume that the sanding result in a workhardened surface layer.)

\vskip 0.3cm
{\bf 3 Plastically deformed surface roughness power spectrum} 

Assume that two elastoplastic solids are squeezed together with a pressure $\sigma_0$
which result in some plastic flow. If the external squeezing pressure is removed the
surface profile is modified. If the external pressure is applied exactly the same way again
the solids are assumed to deform elastically, as is a good approximation in most cases, and the
(elastic) contact area will be the sum $A_{\rm el}+A_{\rm pl}$ of the elastic and plastic contact area
obtained during the first loading act. 
We define the plastically modified power spectrum $C_{\rm pl} (q)$ so that it gives the elastic contact area 
$A=A_{\rm el}+A_{\rm pl}$ for all magnifications $\zeta$.

In earlier publications we have suggested two ways to obtain $C_{\rm pl} (q)$ from the original
power spectrum $C(q)$. Here I present another method which gives {\it exactly} the elastic contact area
$A_{\rm el}+A_{\rm pl}$ when loaded the same way as in the first loading act where the 
surface deformed plastically. We will only focus on contact pressures so that $A/A_0 \lesssim 0.3$.
For such pressures the contact area is approximately linearly related to the the contact pressure $\sigma_0$. 

When the relative contact area $A(\zeta )/A_0$ at the magnification $\zeta = q/q_0$ is less that
$< 0.3$ the contact area is proportional to the squeezing pressure $\sigma_0$:
$${A(\zeta )\over A_0} \approx {\kappa \sigma_0 \over E^* \xi (\zeta)} ,\eqno(6)$$
where $\xi (\zeta)$ is the surface rms-slope including the roughness components with 
wavenumber $q < q_0 \zeta$, and where $\kappa= \surd (8/\pi )$.
We have
$$\xi^2 (\zeta) = 2 \pi \int_{q_0}^{q_0 \zeta} dq q^3 C(q)\eqno(7)$$
Using (1) we get
$$\xi^2 \approx \left [{\kappa \sigma_0 A_0 \over E^* A}\right ]^2 = Q^2(\zeta) .\eqno(8)$$
Using (7) and (8) we get
$$2 \pi \int_{q_0}^{q_0 \zeta} dq q^3 C(q) \approx Q^2 (\zeta)\eqno(9)$$
Taking the derivative of (9) with respect to $\zeta$ gives
$$2 \pi q_0 q^3 C(q) \approx 2 Q(\zeta) Q'(\zeta)$$
or
$$C(q) \approx {\pi \over q_0 q^3} Q(\zeta ) Q'(\zeta)\eqno(10)$$
We can apply this equation to determine $C_{\rm pl}(q)$ if we use $A(\zeta) = A_{\rm el} (\zeta) + A_{\rm pl} (\zeta)$. 
The derivative of $Q'(\zeta)$ is easy to evaluate numerically, and $C_{\rm pl}(q)$ is obtained from (10).

Using $C_{\rm pl}(q)$ in the elastic contact mechanics calculation result in the elastic contact area $A = A_{\rm el}+A_{\rm pl}$.
If $C_{\rm pl}(q)$ is used in the elastoplastic contact mechanics theory then the plastic contact area vanish. This reflect the
assumption described above: If the applied pressure is first increased to $\sigma_0$ the surface deform plastically. Next, if the contact
pressure is removed and then increased again to $\sigma_0$ only elastic deformations occur during this the second loading act.

We will now illustrate the theory with results for the case studied above. The red curve in Fig. \ref{1logq.2logC.plastic.eps}
shows the calculated power spectrum $C_{\rm pl}$ using (10). 
Using this power spectrum we can calculate the elastic contact area $A(\zeta )$, 
given by the red curve in Fig. \ref{1logZeta.2LogArea.all.eps}, which is equal
equals the sum of $A_{\rm el}+A_{\rm pl}$ (sum of the green and blue curves).

\begin{figure}
\includegraphics[width=0.48\textwidth,angle=0.0]{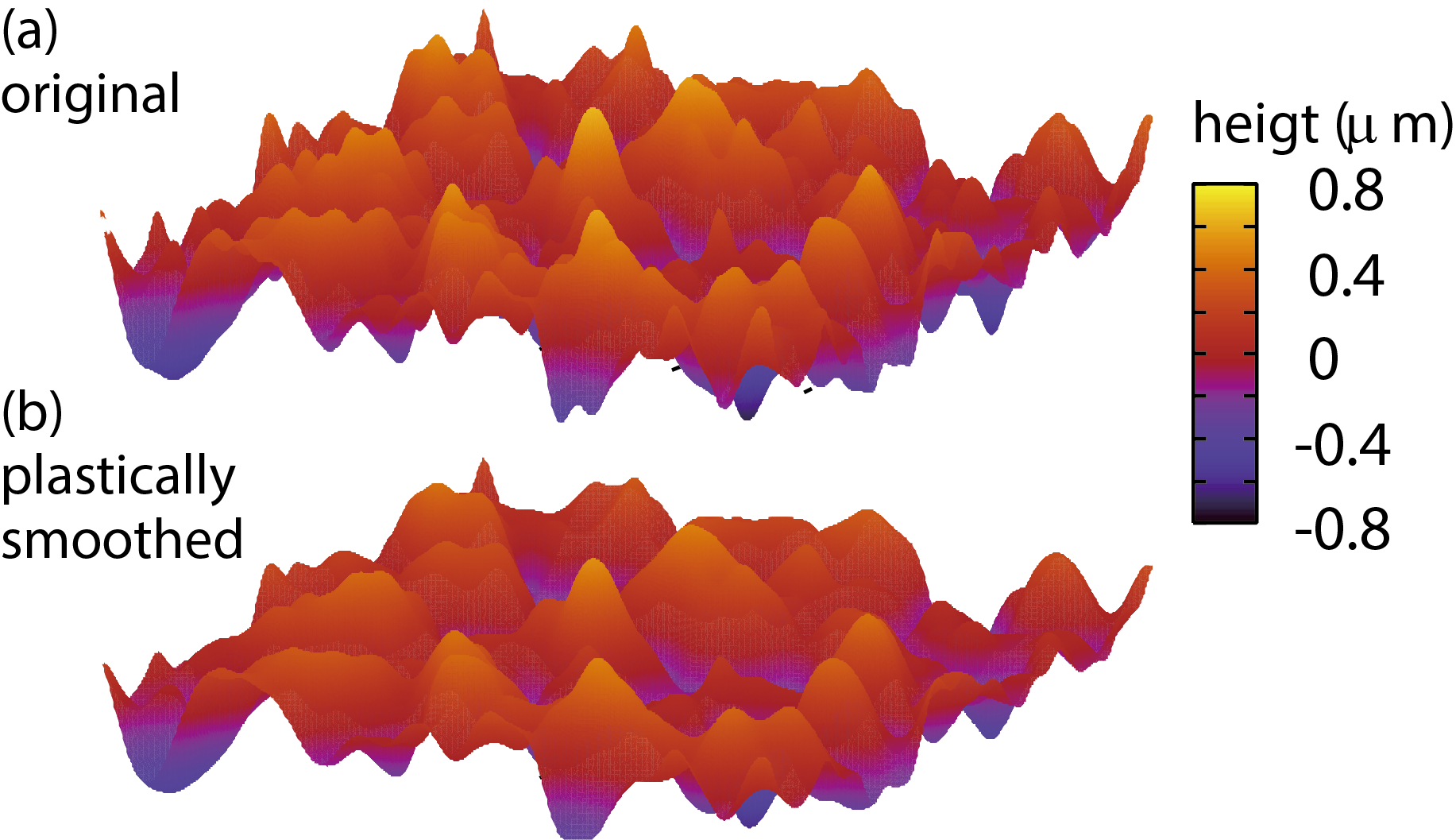}
\caption{\label{Smoothed.original.eps}
The height topography of the original surface (a) and of the plastically smoothed surface (b).
}
\end{figure}

\vskip 0.3cm
{\bf 4 Generating plastically deformed surface height profiles} 

\begin{figure}
\includegraphics[width=0.48\textwidth,angle=0.0]{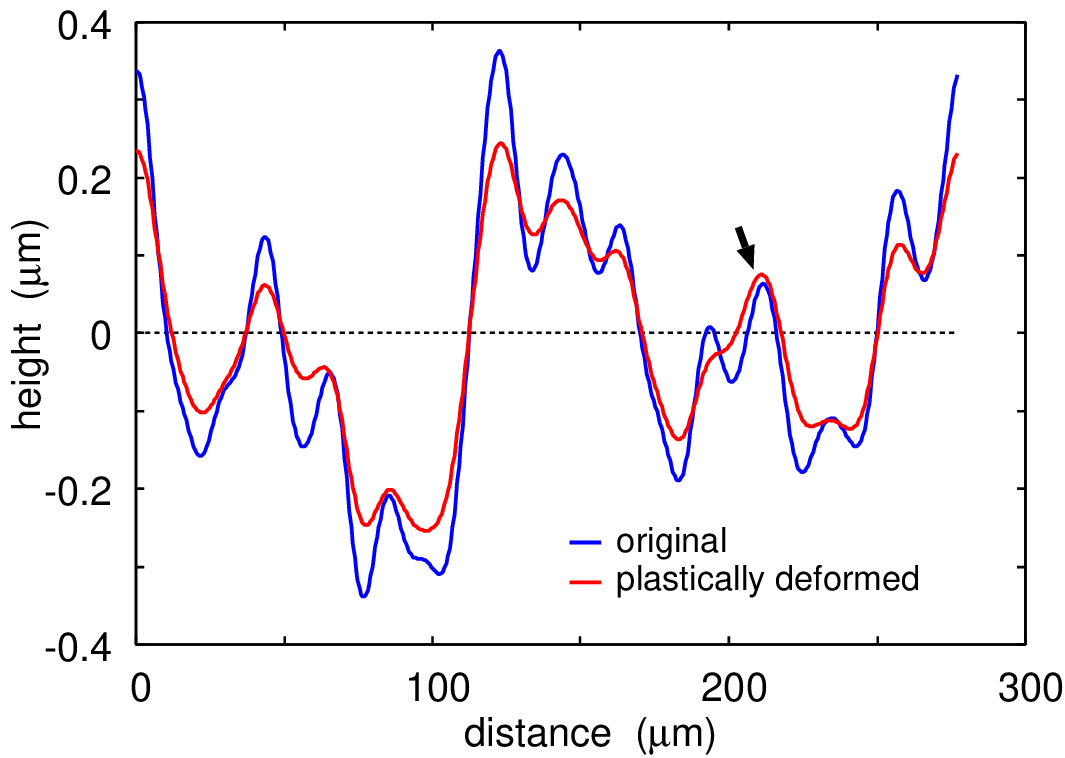}
\caption{\label{1x.2height.1.eps}
The calculated surface height along a line in the 2D surface topography of the original surface (blue line) and
the plastically deformed surface (red line).  Both surfaces have been generated by adding plane waves with random
phases with the weight of the wave components determined by the original and the plastically smoothed power spectra
[given by (10)]. In the two calculations we have used the same set of random numbers. 
}
\end{figure}

Randomly rough surfaces can be obtained by adding plane waves with random phases using
$$h({\bf x}) = \sum_{\bf q} B_{\rm q} e^{i({\bf q} \cdot {\bf x} + \phi_{\rm k})}\eqno(11)$$
where $\phi_{\rm k}$ are random phases uniformly distributed between $0$ and $2\pi$ and
$$B_{\rm k} = {2\pi \over L} [C(q)]^{1/2}\eqno(12)$$
where $L$ is the length of the side of the square area occupied by the topography map.
Using this approach we have generated the 2D surface topography for the original surface using
$C(q)$ (blue line in Fig. \ref{1logq.2logC.plastic.eps}), and for the plastically smoothed surface 
using $C_{\rm pl}$ (red curve in Fig. \ref{1logq.2logC.plastic.eps}), see Fig. \ref{Smoothed.original.eps}. 
In both calculation we use the same set of random numbers.
Fig. \ref{1x.2height.1.eps} shows the height along a line for the original surface (blue line) and for the plastically smoothed surface (red line). 
Note that solids are removed from the highest asperities and moved to the valleys. However, in some cases an asperity below the highest asperities 
becomes higher as indicated by the thick black arrow. 

We note that the smoothing of the profile in the valley in a similar way as at the tops is needed in order for the
plastically smoothed surface to be randomly rough, which is a requirement for the Persson contact mechanics theory. Still, this type of smoothing appears
very similar to observed for steel squeezed against steel in Ref. \cite{exp} and for polymers against flat glass in Ref. \cite{Tiwari2} (see below). 

We will now compare the topography with the measured
results for a rough steel surface squeezed against a very smooth steel surface.
  
\begin{figure}
\includegraphics[width=0.48\textwidth,angle=0.0]{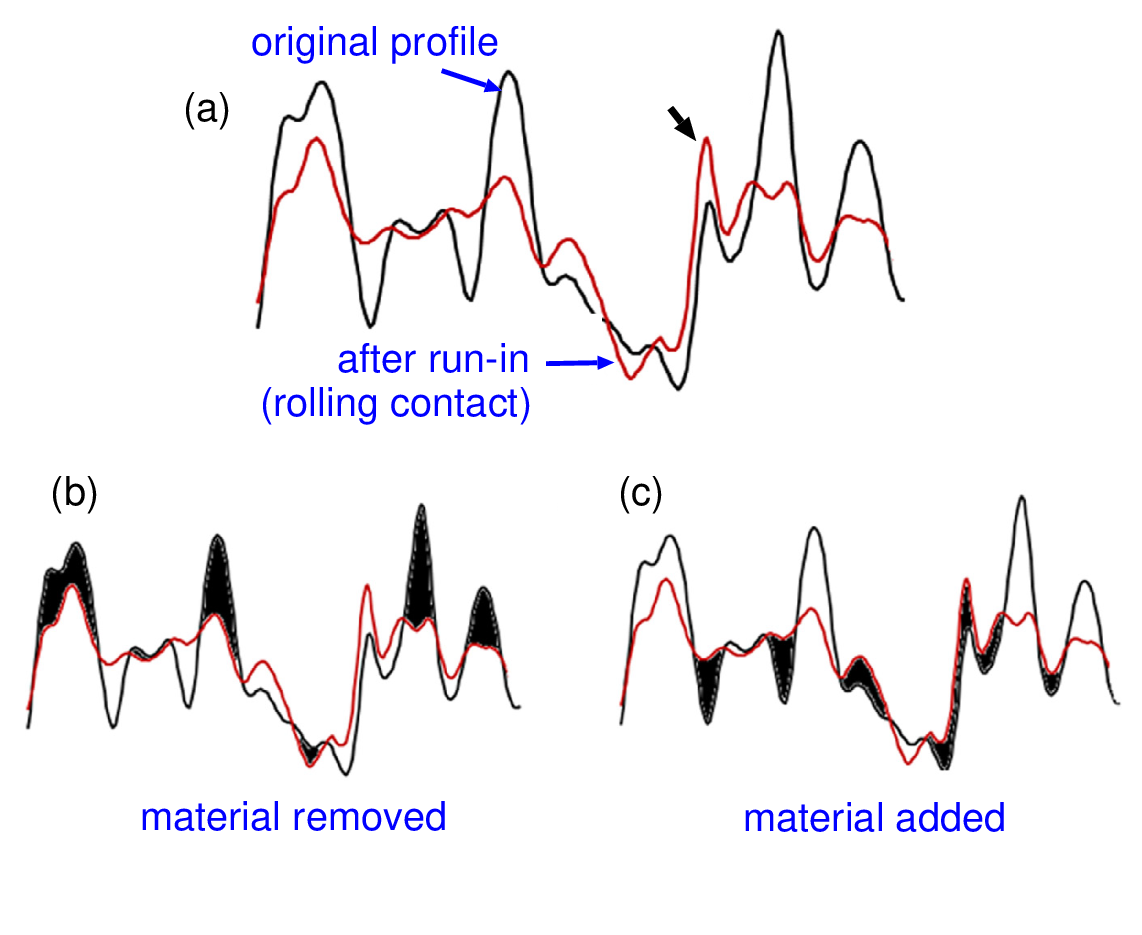}
\caption{\label{RemovedAndDisplacedAndWearMaterial.eps}
(a) The measured height topography along a line on a rough surface before (black line) and after plastic
contact (red line) with another smooth steel surface. (b) and (c) indicate the removed and added material.
Of the removed material $\approx 66\%$ is transferred to the valleys while the rest is attributed to 
material loss (wear particles or metal transfer to the countersurface).
}
\end{figure}

\vskip 0.3cm
{\bf 5 Comparison with experiment} 

The change in surface topography caused by plastic deformation has been studied by Yusof and Ripin 
in Ref. \cite{exp}. The experiment consisted of a steel cylinder with very smooth surface rolling against another
steel cylinder with much larger roughness prepared using sandpaper. 
The smooth surface was sanded by 1500 grit sandpaper, 
and rough with 120 grit sandpaper, resulting in $\sim 50$ times larger roughness amplitude on the rough cylinder surface.
Both cylinders have the radius $R=1 \ {\rm cm}$.
The cylinders where squeezed in contact so that a rectangular Hertz contact formed with the width (in the rolling direction)
$51 \ {\rm \mu m}$ and with maximum Hertz contact pressure $p= 0.209 \ {\rm GPa}$. From contact mechanics one expect
that a shear stress develop in the nominal contact area, but this is much smaller than the
normal stress in most of the nominal contact area, but may influence the formation of wear particles. 

Fig. \ref{RemovedAndDisplacedAndWearMaterial.eps} shows the measured height profile along a $\sim 250 \ {\rm \mu m}$ 
long track before (black line) and after 10 rolling cycles (red line). 
Fig \ref{RemovedAndDisplacedAndWearMaterial.eps}(b) and (c) indicate the removed and added material.
Of the removed material $\approx 66\%$ is transferred to the valleys while the rest is attributed to 
material loss (wear particles or metal transfer to the countersurface). In some locations asperities below the
highest asperities become higher after the plastic deformations, indicated by the thick black arrow in one case, 
as also observed in the simulations.

The experimental measured height profile is statistically very similar to the simulated profile with most of the displaced
material ending up in the valley. Since some part of the material removed from the tops of the asperities end up as wear particles,
or attached to the smooth countersurface, a detailed comparison with the simulations would not be possible even if the surface
roughness powerspectrum of the original surfaces would be known.

\begin{figure}
\includegraphics[width=0.48\textwidth,angle=0.0]{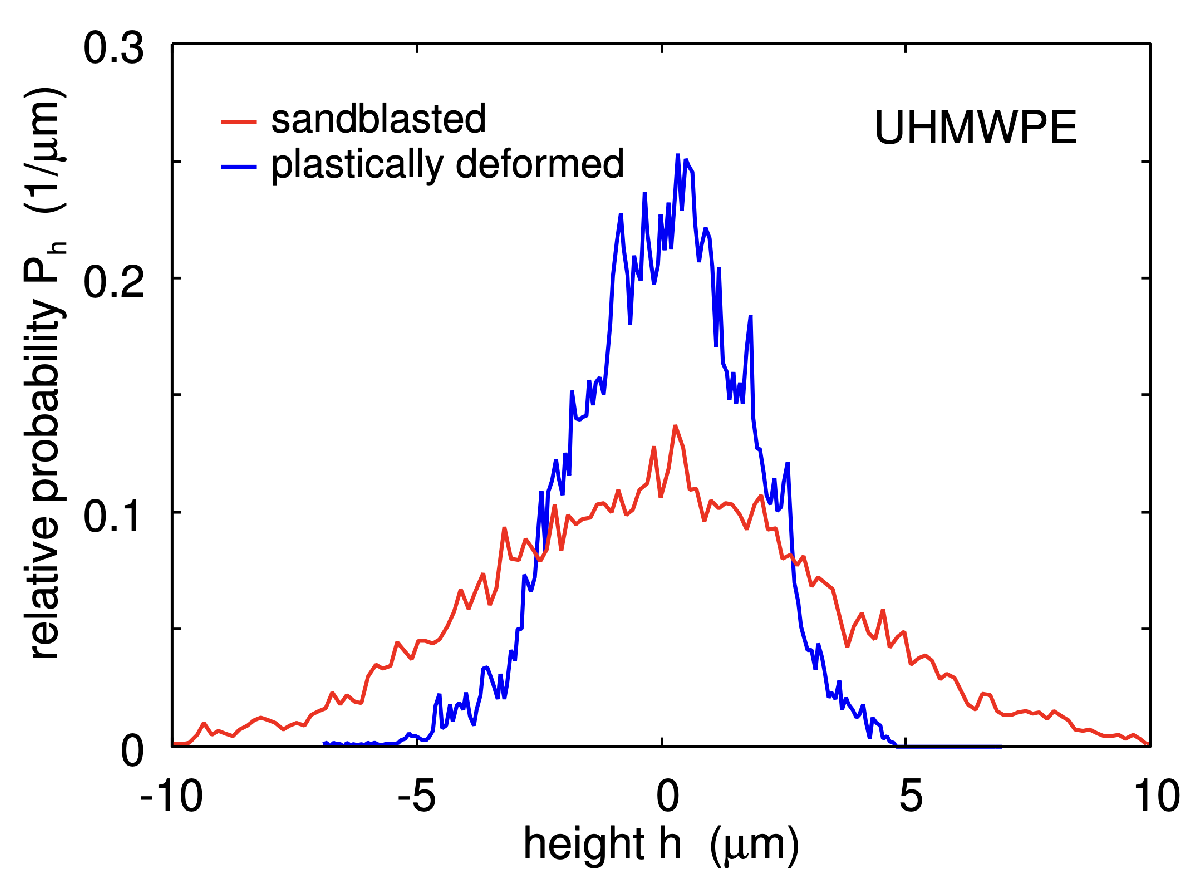}
\caption{\label{Poly.eps}
The surface height probability distribution as a function of the surface height for a sandblasted
polyethylene surface (red) and the plastically deformed surface obtained by 
squeezing the polymer against a flat glass surface (blue).
Adapted from \cite{Tiwari2}.
}
\end{figure}

\begin{figure}
\includegraphics[width=0.35\textwidth,angle=0.0]{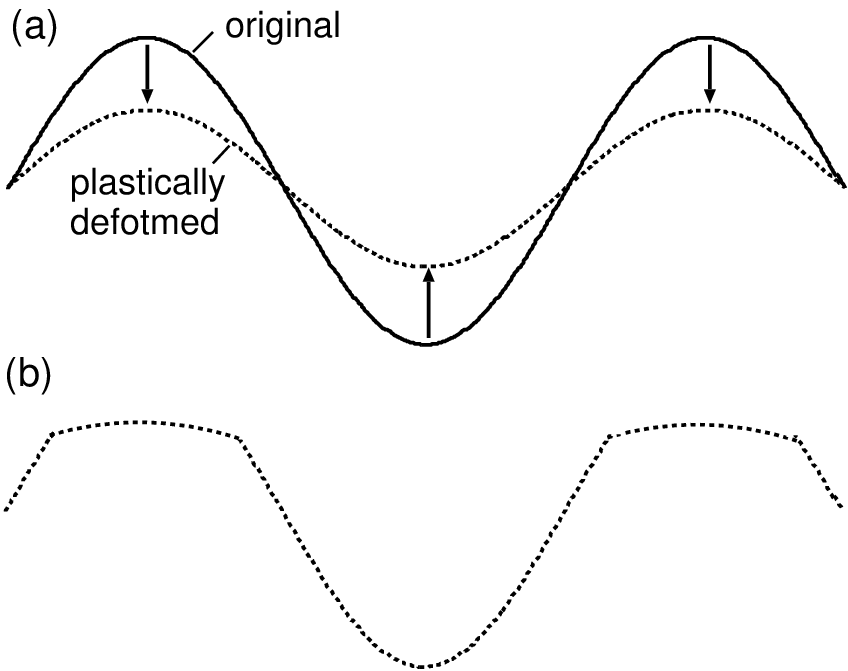}
\caption{\label{1x.2sinus.eps}
(a) A sinus roughness deform by plastic flow into another sinus roughness
with smaller amplitude. (b) A sines roughness deform to form a flattened top
surface area. We refer to the smoothing in (a) and (b) as mode-1 and mode-2
smoothing.
}
\end{figure}

\begin{figure}
\includegraphics[width=0.48\textwidth,angle=0.0]{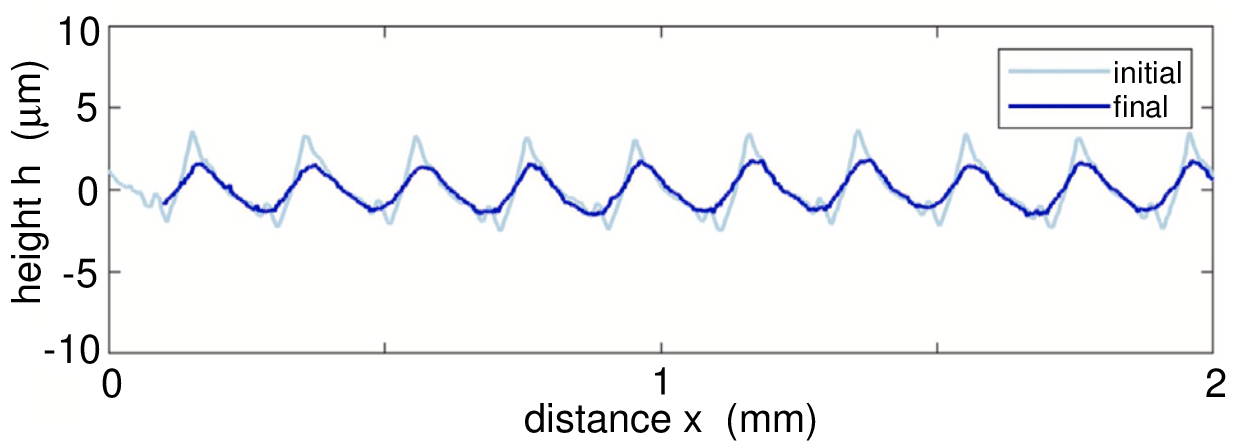}
\caption{\label{Another.x.h.eps}
The height profile on a turned surface with a periodic height profile.
After rolling contact both the tops and valleys get smoothed bu plastic flow.
Adapted from Ref. \cite{periodic}.
}
\end{figure}

\vskip 0.3cm
{\bf 6 Discussion} 

There is one mystery in plastic deformation of rough surfaces which I will discuss here. 
In an earlier publication we have studied the change in the surface topography of polymer
surfaces after they where squeezed in contact with very a smooth glass surface \cite{Tiwari2}. The polymer surfaces were sandblasted 
and had nearly Gaussian height profiles with the roughness below the average surface plane appearing statistically the same as above. 
After plastic deformations the polymer surfaces had still Gaussian 
roughness but with reduced width, and the surface topography still appeared just as randomly rough as before the plastic deformation. 
This was observed for polycarbonate, nylon, polyethylene, polypropylene,
and polytetraflourethylene and is illustrated in Fig. \ref{Poly.eps} for polyethylene. We suggested in 
Ref. \cite{Tiwari2} that this is due to flow of material from the top of asperities to the valley in such a way
that a sinus corrugation component in the roughness profile remains as a sinus component but with reduced
amplitude as indicated in Fig. \ref{1x.2sinus.eps}.

Another illustration of the influence of plastic flow on the surface topography is shown in 
Fig. \ref{Another.x.h.eps}. In this case a steel cylinder with a periodic height profile (turned surface)
was rolled against another cylinder with smoother surface. The Hertz nominal contact pressure was very high
($\sim 1-2 \ {\rm GPa}$), but BEM calculations indicated that the contact pressure in the valleys is well below
the steel penetration hardness\cite{periodic}. Nevertheless, the topography in Fig. \ref{Another.x.h.eps} shows that 
the rolling result in plastic flow which smooth both the tops and valleys.  

\begin{figure}
\includegraphics[width=0.48\textwidth,angle=0.0]{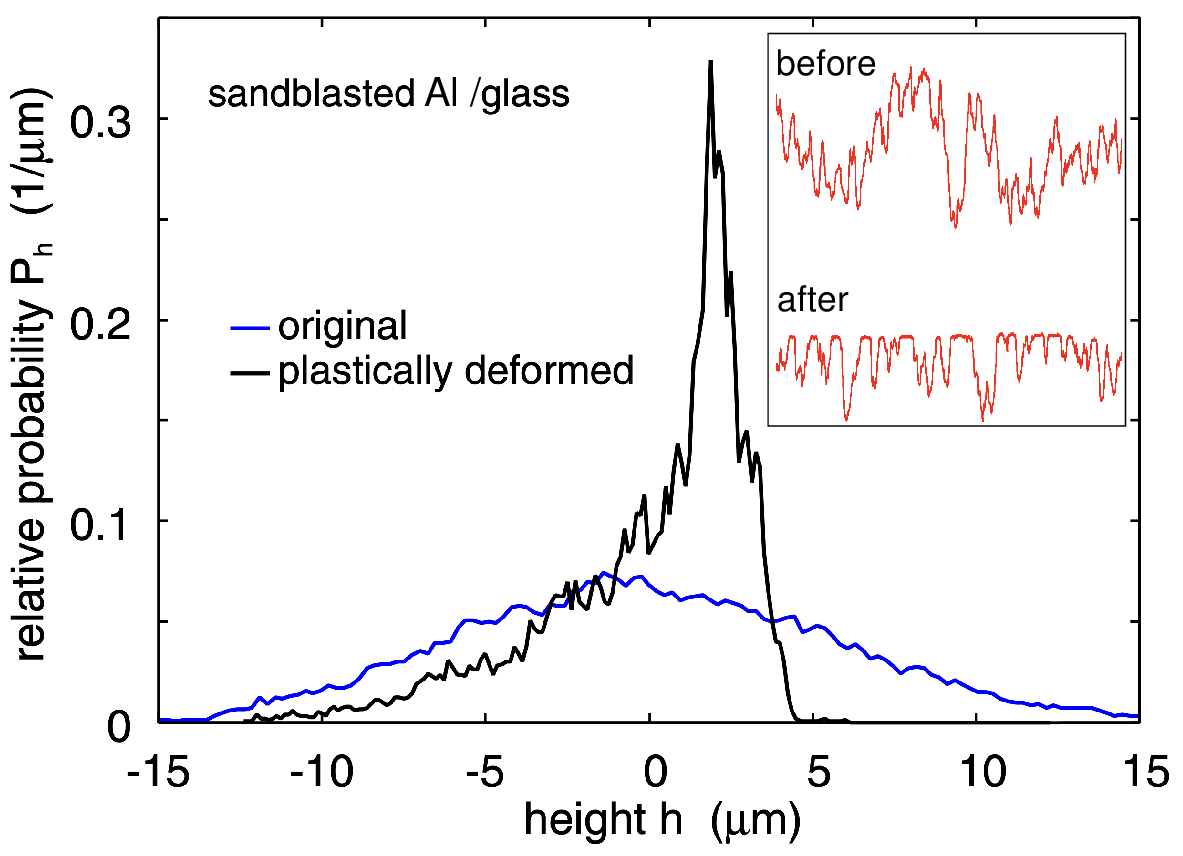}
\caption{\label{PlastAlWithInset.eps}
The surface height probability distribution as a function of the surface height for a sandblasted
aluminum (blue) and the plastically deformed surface obtained by 
squeezing the aluminum against a flat steel surface (black). The inset shows two lines scans
(from different locations) on the aluminum surface before and after the plastic deformation.
Adapted from \cite{Tiwari}.
}
\end{figure}

\begin{figure}
\includegraphics[width=0.48\textwidth,angle=0.0]{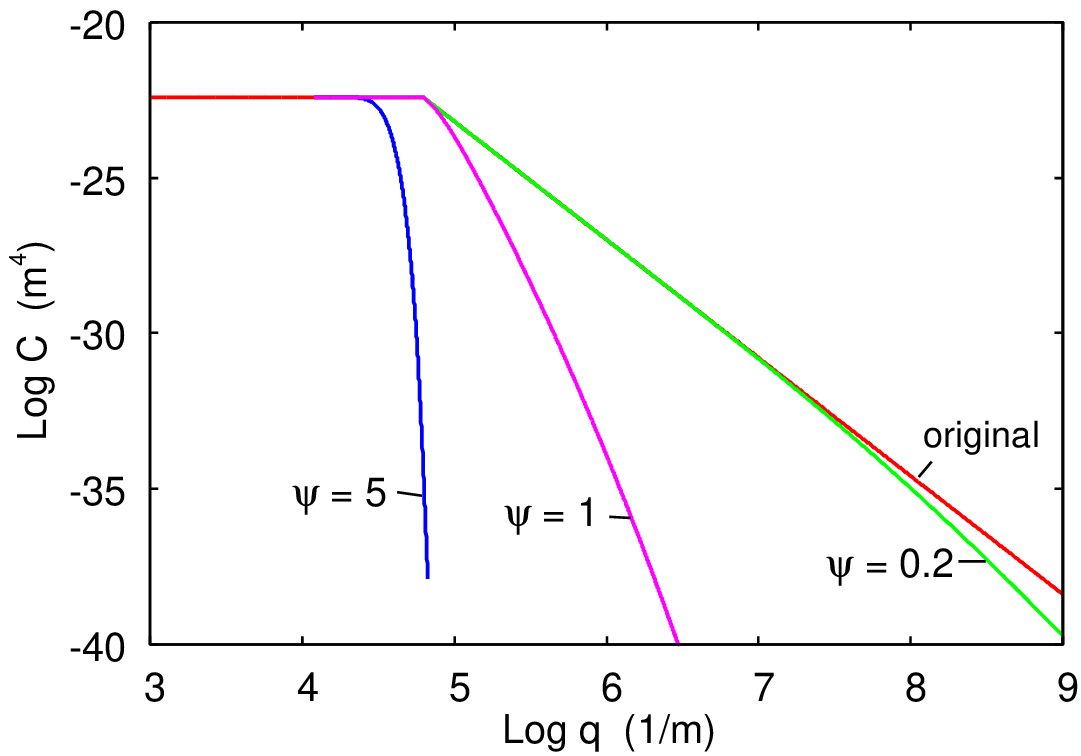}
\caption{\label{1logq.2logC.many.psi.eps}
The plastically smoothed power spectra for several values of the plasticity index. 
}
\end{figure}

For the steel surface studied by Yusof and Ripin no height distribution was presented but the 
given topography pictures shows similar roughness above and below the average surface plane
also after the plastic deformation. This result differ strongly from measurements for sandblasted
aluminum squeezed against flat countersurfaces \cite{Williams,Tiwari}, where the top of the asperities flatten and where the
height probability distribution becomes highly asymmetric as illustrated in Fig. \ref{PlastAlWithInset.eps}.
This may relate to different plastic flow properties of the polymers and the steel compared to aluminum.
Here I will propose a different explanation. The nature of the elastoplastic contact is characterized 
by the plasticity index (see Ref. \cite{PRLcomment}): 
$$\psi = {1\over 2} q_{\rm r} h_{\rm rms} {E^*\over \sigma_{\rm P}} ,\eqno(13)$$
where $q_{\rm r}$ is the roll-off wavenumber of the surface roughness power spectrum.
Fig. \ref{1logq.2logC.many.psi.eps} shows for a typical case 
the plastically smoothed power spectra for several values of the plasticity index. 
I have found that for plastically smoothed surfaces with $\psi \approx 1$ the plastic smoothing
start at the onset of the roll-off region as in Fig. \ref{1logq.2logC.many.psi.eps} (pink line).
This indicate a qualitative change in the smoothing when $\psi$ changes from $<1$ to $>1$. I propose that for $\psi \lesssim 1$
the smoothing occur as in Fig. \ref{1x.2sinus.eps}(a) and as in Fig. \ref{1x.2sinus.eps}(b) for $\psi \gg 1$.
I will refer to these as mode-1 and mode-2 smoothing, respectively.

Using literature values or measured values for the penetration hardness $\sigma_{\rm P}$, the effective modulus
$E^*$ and the measured rms-roughness $h_{\rm rms}$ and roll-off wavenumber $q_{\rm r}$ \cite{explain1}
we estimate $\psi \approx 0.2$ for the steel contact studied 
in Ref. \cite{exp} (see Fig. \ref{RemovedAndDisplacedAndWearMaterial.eps}), and $\psi \approx 0.9$ the 
PE polymer surface in Ref. \cite{Tiwari2}. For the aluminum surfaces used in the study in  Ref.
\cite{Tiwari} I estimate $\psi \approx 8-23$. This support the proposal that the different mode of plastic flow may reflect the magnitude
of the plasticity index. If $\psi \lesssim 1$ the plastic flow is as indicated in Fig. \ref{1x.2sinus.eps}(a) while if 
$\psi \gg 1$ then the flow is more like in Fig. \ref{1x.2sinus.eps}(b). To lend additional support for this proposal I note that
the study in Ref. \cite{Inose} for copper and steel shows similar mode-2 plastically deformed profiles as for 
aluminum in Fig. \ref{PlastAlWithInset.eps} 
(Ref. \cite{Tiwari}) and for these cases one can estimate $\psi \approx 10$ and $ \approx 6$, respectively.
In table \ref{Psiindex} we give the parameters used in the estimations of the plasticity index.

%
\begin{table}[hbt]
   \caption{The plasticity index for several systems. The references are given in the square brackets [..].
The experiments show that the systems with $\psi < 1$ exhibit mode-1 smoothing and those with $\psi > 1$ mode-2 smoothing.}
   \label{Psiindex}
   \renewcommand{\arraystretch}{1.5} 
   \begin{center}
      \begin{tabular}{@{}|l||c|c|c|c|c|@{}}
         \hline
            system & $q_{\rm r} \ ({\rm 1 /\mu m})$ & $h_{\rm rms} \ ({\rm \mu m})$ & $E^* \ ({\rm GPa})$ & $\sigma_{\rm P} \ ({\rm GPa})$ & $\psi$ \\
         \hline
         \hline
            Al [22] & 0.03  & 16.0 & 75 & 0.8 & 23 \\ 
         \hline
            Al [22] & 0.03  & 6.0 & 75 & 0.8 & 8 \\ 
         \hline
            PE [20] & 0.05 & 4.0  & 0.7 & 0.075 & 0.9 \\
         \hline
            steel [6] & 0.03 & 0.5  & 100 & 3.2 & 0.2 \\
         \hline
            steel [29] & 0.1 & 3.0 & 113 & 1.8 & 10 \\
         \hline
            Cu [29] & 0.1 & 1.4 & 84 & 0.94 & 6 \\
         \hline
      \end{tabular}
   \end{center}
\end{table}
%

This transition from flow-to-the-valley for $\psi \lesssim 1$ 
to some other form of plastic flow for $\psi \gg 1$ may result from plasticity mechanics 
of asperity interaction \cite{A23,A24,A25,A26,A27,A28,Inose,Xu}.
As the asperities coalesce and merge the hydrostatic stress dominate. 
As a result, the material will plastically deform initially 
but then revert to elastic deformation, which is believed to be the  origin of why the contact area saturate 
at $A/A_0 \approx 0.5$ for large nominal contact pressure.

\vskip 0.3cm
{\bf 7 Summary and conclusion} 

When two metal blocks are squeezed together the stresses in the asperity contact regions are usually so large that
the asperities deform plastically, at least at short length scales. 
Many tribology properties of contacts, such as the contact stiffness and the electric and thermal contact resistance, and the fluid flow at interfaces,
depend on the surface topography and are hence modified by the plastic flow. I have present a new way to obtain an effective power spectra of
plastically deformed surfaces to be used in the Persson contact mechanics theory. I have also present results for the
surface height topography obtained using the plastically modified power spectra,
and compared to the experimental results of Yusof and Ripin, who studied the influence of plastic flow on the 
topography for a smooth steel surface squeezed against a rough steel surface. Finally, I have discussed 
why some surfaces after plastic deformation have similar Gaussian roughness as before plastic deformation, only with smaller roughness amplitude,
while other surfaces shows very skewed roughness after plastic deformation.

\vskip 0.3cm

{\bf Acknowledgments}

I thank Robert Jackson for comment on the text.


\begin{thebibliography}{}
\bibitem{explain}
When I state ``plastic flow from top of asperities to the valleys'' I do not mean that the solid
at the top of asperities move to the valleys but that when the solid moves downwards at the top of asperities 
it simultaneously moves upwards in the valleys by the same amount so that the volume is conserved.
\bibitem{PRLcomment}
The Persson contact mechanics theory \cite{PRL} shows that for elastoplastic contact
for a self-affine fractal surface with a roll-off the contact area depends on the Hurst exponent $H$, the normalized squeezing pressure
$\sigma_0/\sigma_{\rm P}$ and the plasticity index given by (13). Another plasticity index was introduced by
Greenwood and Williamson \cite{Green} but this depends on an asperity radius which is ill-defined for surfaces with
multiscale roughness.
\bibitem{explain1}
The roll-off wavenumber can be obtained from the surface roughness power spectrum
$C(q)$ or, when $C(q)$ is not measured, from visual inspection of topography line scans.
\end{thebibliography}
\end{document}